\documentclass[prd,twocolumn,showpacs,preprintnumbers,superscriptaddress,nofootinbib,amsmath,amssymb,nobalancelastpage,floatfix]{revtex4}
\usepackage{graphicx}
\usepackage{bm}
\usepackage{dcolumn}
\usepackage{amsmath}
\usepackage{amssymb}
\usepackage{color}
\usepackage{url}
\usepackage{aas_macros}

\usepackage{hyperref}
\usepackage{subfigure}

\usepackage{color}

\newcommand\rhocrit{\rho_{\mathrm{c}}}
\newcommand\Omegam{\Omega_{\mathrm{m}}}

\begin{document}

\title{Cosmological constraints to dark matter with two- and many-body decays}

\author{Gordon Blackadder}
 \email{blackadder@brown.edu}
\author{Savvas M. Koushiappas}
\email{koushiappas@brown.edu}
\affiliation{Department of Physics, Brown University, 182 Hope St., Providence, RI 02912}

\date{\today}

\begin{abstract}
We present a study of cosmological implications of generic dark matter decays. We consider two-body and many-body decaying scenarios. In the two-body case the massive particle has a possibly relativistic kick velocity and thus possesses a dynamical equation of state. This has implications to the expansion history of the universe. We use recent observational data from the cosmic microwave background, baryon acoustic oscillations and supernovae Type Ia to obtain constraints on the lifetime of the dark matter particle. We find that for an energy splitting where more than 40\% of the dark matter particle energy is transferred to massless, relativistic particles in the two-body case, or more than 50\% in the many-body case, lifetimes less than the age of the universe are excluded at more than 95\% confidence. When the energy splitting falls to 10\% the lifetime is constrained to be more than roughly half the age.
\end{abstract}

\maketitle

\section{\label{sec:level1}Introduction}

The presence of dark matter in the universe necessitates a fundamental mechanism of its generation. In most dark matter scenarios, the present-day dark matter is the lowest quantum state of a beyond the Standard Model theory, whose longevity is guaranteed by some new global symmetry of the theory. As such, in most dark matter theories the dark matter particle is the product of a decay scenario, and therefore  in all of these cases the main question is how and when this decay takes place in the cosmological history of the universe \cite{Berezhiani:2015aa,Enqvist:2015aa,Esmaili:2014aa,Geng:2015aa,Hamaguchi:2015aa,Lovell:2014aa,Iakubovskyi:2014aa,Ando:2015aa,Audren:2014aa,Perez-Garcia:2015aa,Rott:2014aa,Yang:2015aa, Blackadder:2014aa}. 

The motivation for decaying dark matter is multi-fold: it arises in purely theoretical considerations, phenomenological models of experimental results, as well as cosmological arguments regarding small scale structure in the universe. 
An example of a purely theoretical motivation is in supersymmetric theories, where the dark matter particle today is the product of the decay of the second-to-lightest particle, that being the gravitino \cite{Bomark:2014aa}, the gaugino\cite{Ibarra:2009aa} or the sneutrino\cite{Asaka:2007aa} (see \citet{Ibarra:2013aa} for a full review). Similar arguments can be made to theories with universal extra dimensions \cite{Hooper:2007aa}.

An example of phenomenological modeling of experimental results in the context of decaying particle is in the context of the recent results from IceCube on the South Pole \cite{Aartsen:2014aa} and the AMS-02 experiment onboard the International Space Station \cite{Accardo:2014aa,Aguilar:2014aa}. The hypothesis in these cases is that these unexpected observations of very high energy neutrinos and/or a high positron fraction with a cutoff can be explained by the decay of a dark matter candidate (regarding the high energy neutrinos see  \cite{Anchordoqui:2014aa} and see \cite{Geng:2015aa,Hamaguchi:2015aa} and references therein for the AMS-02 positron fraction). Another experimentally interesting measurement is the presence of an excess of 3.5 keV photons along the line of sight to the Andromeda galaxy (M31) as well as other galaxies  and galaxy clusters (as observed in a stacked analysis) \cite{Boyarsky:2014aa,Bulbul:2014aa}. While at the moment there is no obvious or generally accepted astrophysical origin, such a measurement can  be the result of the daughter products of decaying dark matter \cite{Lovell:2014aa,Iakubovskyi:2014aa} (the latter of which is a review article that also looks at possible astrophysical explanations).

In addition, large scale structure considerations perhaps point to the need for a dark matter particle with non-zero thermal velocity. Such an effect can be accomplished by dark matter decay, where the present-day dark matter particle possesses a ``kick" velocity as a by-product of an earlier decay from a parent heavy particle (the mass difference between the parent particle and one of the daughters is very slight but due to energy and momentum conservation the most massive daughter has a higher velocity than the parent -- see e.g., \cite{Peter:2010aa,Peter:2010ab,Wang:2013aa,Wang:2014aa,Cheng:2015aa}). 

From the examples above we see that  decays are an intrinsic part of the cosmological history of the universe motivated by both theoretical and experimental grounds. The goal then is to explore the mechanics of such processes in a cosmological context. 

In a previous study \cite{Blackadder:2014aa}, we developed and examined in detail the physics of a most generic model of decaying dark matter in which the decay proceeds either to two- or many-body final states as the universe expands. One of the main features of that work was the derivation of the equation of state of the daughter particles, thus allowing the daughter particles (dark matter or not) to dynamically affect the rate of expansion of the universe. We then explored the constraints to such a generic decay scenario imposed by the expansion history of the late universe using type Ia supernova (SNe) observations\cite{Suzuki:2012aa}. 

In this paper we expand significantly on the analysis presented in \cite{Blackadder:2014aa}. We consider many- and two-body decaying dark matter using a Markov Chain Monte Carlo analysis that constrains against distance measurements from Planck and WMAP CMB measurements \cite{Planck-Collaboration:2014aa,Hinshaw:2013aa}, Galaxy Baryon Acoustic Observations (BAO) \cite{Anderson:2014aa,Beutler:2011aa,Ross:2015aa} and Lyman-$\alpha$ Forest BAO \cite{Delubac:2015aa,Font-Ribera:2014aa} as well as the Joint Light-Curve Analysis of 740 type Ia Supernovae \cite{Betoule:2014aa}. 
In this expanded analysis the principle remains the same. While there are many specific models of decaying dark matter, which due to their very specific nature can sometimes be constrained quite highly, it is useful to consider what constraints can be put on a generic model that will then have wide applications and implications. Considering decays into a massless, relativistic component and possibly into a massive component which may or may not have kinetic energy, the expansion history of the universe is changed in a manner that can be constrained by distance observations. 

The paper is organized as follows: Section \ref{sec:level2} reviews the models of decaying dark matter. In Section \ref{sec:level3}  we discuss in detail the improvements between the previous and present analyses, while in Section \ref{sec:level4} we present the results including constraints on the parameters that were allowed to vary in the analysis ($\Omegam$, $\Omega_{b}h^{2}$, $h$ and the decay rate $\Gamma$) along with confidence limits on the decaying dark matter lifetime. We conclude in section \ref{sec:level5} with a review of the recent literature.

\section{\label{sec:level2}Decaying dark matter in an expanding universe}

In this section we will briefly summarize the decaying dark matter model of \cite{Blackadder:2014aa}, which forms the basis of the analysis carried out here. 

This model assumes that there is a parent particle, assumed to be at rest, which decays exponentially with lifetime $\tau=1/\Gamma$. We  consider two scenarios: First a two-body decay in which the daughter products are a single massless, relativistic daughter and a single massive daughter. The possibly relativistic velocity of this massive daughter is  set by the requirements of energy and momentum conservation but the particle is  then allowed to slow with the expanding universe. Secondly we consider a many-body decay in which there are many massless, relativistic daughters as well as a single massive daughter. In this scenario a particular kick velocity for the massive daughter cannot be determined and so it is assumed to be stationary (see \cite{Blackadder:2014aa} for more details).

The parent, the massless duaghter(s) and massive daughter particles are labeled 0, 1 and 2 respectively. We define $\epsilon$ as the fraction of the parent particles' (of mass $m_0$) energy that is transferred to the {\em massless} daughter particle(s), the remainder being transferred to the {\em massive} daughter (of mass $m_2$). In the case of the many-body decay, $\epsilon$ is related to the two masses via the trivial relation

\begin{equation}
\epsilon_{\mathrm{many}} = \frac{m_{0}-m_{2}}{m_{0}}
\end{equation}
while in the case of two body decay the relationship is

\begin{equation}
\epsilon_{\mathrm{two}} = \frac{1}{2}\left(1-\frac{m_{2}^{2}}{m_{0}^{2}}\right)
\end{equation}

In both, two- and many-body decays, the parent particle density and the massless daughter density obey

\begin{eqnarray}
\frac{d\rho_{0}}{dt} + 3\frac{\dot{a}}{a}\rho_{0} &=& -\Gamma \rho_{0}\\
\frac{d\rho_{1}}{dt} + 4\frac{\dot{a}}{a}\rho_{1} &=& \epsilon\Gamma\rho_{0}
\end{eqnarray}
respectively. 

In the many-body model the massive daughter evolves according to the equation

\begin{equation}
\frac{d\rho_{2, \mathrm{many}}}{dt} + 3\frac{\dot a}{a}\rho_{2, \mathrm{many}} = (1-\epsilon)\Gamma\rho_{0}
\end{equation}¥

Such a straight forward equation cannot be written in the two-body case owing to the equation of state which changes with time in a non-trivial manner as some particles are created by the decay at a high velocity while at the same time other particles, created at an earlier epoch, slow due to the expansion of the universe. The density can instead be found to obey

\begin{eqnarray}
\rho_{2, \mathrm{two}}=\frac{{\cal{A}} \Gamma\sqrt{1-2\epsilon}}{a^{3}}\int_{a_{\star}}^{a}{\cal{J}}(a,a_{D})da_{D} \nonumber\\
{\cal{J}}(a, a_{D}) \equiv \frac{e^{-\Gamma t(a_{D})}}{a_{D}H_{D}}\sqrt{\frac{\beta_{2}^{2}}{1-\beta_{2}^{2}}\left(\frac{a_{D}}{a}\right)^{2}+1}
\end{eqnarray}
where ${\cal{A}}$ is a constant set by the present day dark matter density, the hubble parameter is $H_{D}=H(a_{D})$ evaluated at some earlier epoch of the expansion parameter $a_{D}$, and $\beta_{2}$ is the initial kick velocity of the heavy daughter determined by $\epsilon$, $\beta_{2}^{2} = \epsilon^{2} /(1 - \epsilon)^{2}$ (the full derivation is contained in \cite{Blackadder:2014aa} including the equation of state for the massive daughter in the two-body scenario). $a_{\star}$ is an arbitrarily small value of the expansion parameter before which, it is assumed, no significant number of decays have occurred.

In the context of this paper it is computationally useful to describe the densities in terms of first order differential equations of a scaled quantity, $r$  \cite{Aubourg:2014aa}, as:

\begin{eqnarray}
\rho_{0}(a) = \rhocrit a^{-3} r_{0}(a)\\
\rho_{1}(a) = \rhocrit a^{-4}r_{1}(a)\\
\rho_{2}(a) = \rhocrit a^{-3}r_{2}(a)
\end{eqnarray}¥ 

For both the two- and the many-body scenarios the evolution of the parent and massless daughter particles can be expressed as

\begin{eqnarray}
\frac{dr_{0}}{d \ln{a}} &=& - \frac{\Gamma \, r_{0}}{H(a)}\label{EQ:Parent}\\
\frac{dr_{1}}{d \ln{a}} &=& \epsilon \frac{ \Gamma \, r_{0}}{H(a)}a \label{EQ:MasslessDaughter}
\end{eqnarray}
where we draw particular attention to the factor of $a$ in Eq. \ref{EQ:MasslessDaughter} that arises due to  cosmological expansion (redshifting). 

In the many-body decay the evolution of the heavy daughter is simply

\begin{equation}
\frac{dr_{2}}{d \ln{a}} = (1- \epsilon) \frac{ \Gamma \, r_{0}}{H(a)}\label{EQ:ManyMassiveDaughter}
\end{equation}¥ 

The derivation of the evolution of the heavy daughter in the two-body decay is a little more involved. We start by looking at the fact that $dn_2(a_D) = - d n_0$, i.e., the change in the number density of the heavy daughter is equal to minus the change in the number density of the parent particles, 
\begin{eqnarray}
\frac{dn_{2}(a_{D})}{dt_{D}} &=& -\frac{dn_{0}}{dt_{D}} \nonumber \\
&=& -\frac{d(\rhocrit r_{0})}{m_{0}dt_{D}}, \nonumber
\end{eqnarray}
which can be rewritten as 

\begin{equation}
\frac{dn_{2}(a_{D})}{d \ln{a}} = \frac{\Gamma}{H(a_{D})} \frac{\rhocrit  r_{0}(a_{D})}{m_{0}}
\label{eq:dn2}
\end{equation}
As the energy density of the heavy daughter is (see \cite{Blackadder:2014aa}), 
\begin{equation}
\rho_{2}(a) = a^{-3}\int_{n(a_{\star})}^{n(a)} E_{2}(a, a_{D}) dn_{2}(a_{D}),
\label{eq:rho2}
\end{equation}
substitution of Eq.~\ref{eq:dn2} in Eq.~\ref{eq:rho2} gives 
\begin{equation}
\rho_{2}(a) = a^{-3}\int_{\ln{a_{\star}}}^{\ln{a}}\frac{E_{2}(a, a_{D})}{m_{0}}\frac{\Gamma \rhocrit r_{0}(a_{D})}{H(a_{D})}d\ln{a_{D}}
\end{equation} 
from which we can readily write the expression of the dimensionless density variable of the heavy daughter as 
\begin{equation}
r_{2}(a) = \int_{\ln{a_{\star}}}^{\ln{a}} \frac{E_{2}(a, a_{D})}{m_{0}}\frac{\Gamma r_{0}(a_{D})}{H(a_{D})}d\ln{a_{D}}. 
\end{equation}
Differentiating with respect to $\ln{a}$ then gives\footnote{This requires  using the Leibniz integral rule on the right hand side.}
\begin{eqnarray}
\frac{dr_{2}(a)}{d\ln{a}} &=& (1-\epsilon)\frac{\Gamma r_{0}(a)}{H(a)}\nonumber \\&&+ \frac{\zeta \sqrt{1-2\epsilon}}{a^{2}}\int_{r_{0}(a_{\star})}^{r_{0}(a)}   {\cal I}(a, a_{D}) d r_{0} 
\end{eqnarray}
where
\begin{equation}
{\cal I}(a, a_{D}) \equiv  \frac{a_{D}^{2}}{\left[\zeta (a_{D}^{2}/a^{2}) + 1\right]^{1/2}} 
\end{equation}
and  $\zeta = \beta_{2}^{2}/(1-\beta_{2}^{2})$.

\section{\label{sec:level3}  Cosmological constraints on decaying dark matter}

To constrain any decaying dark matter scenario it is necessary to choose  boundary conditions; conditions that fix the dark matter density at some epoch in the history of the universe. A natural boundary condition is the abundance of dark matter at the epoch of recombination. This is the approach taken in \cite{Blackadder:2014aa} where initial conditions are set by the Planck 2014 results (a combination of the first Planck data release \cite{Planck-Collaboration:2014aa}, low-l WMAP data \cite{Hinshaw:2013aa} and high-l Atakama Cosmology Telescope \cite{Das:2014aa} and the South Pole Telescope \cite{Reichardt:2012aa} data). The assumption is that these initial conditions are true before any significant decay has occurred (see discussion in Section V, in \cite{Blackadder:2014aa}). Under this assumption, \cite{Blackadder:2014aa} uses the 580 type Ia supernovae of the Union2.1 catalog \cite{Suzuki:2012aa}  to constrain the decay rate $\Gamma$ and energy splitting fraction $\epsilon$ by evaluating a goodness-of-fit based on the sum of $\chi^{2}$ values for each supernova. 

Taking the Planck14 results as fixed initial conditions at the era of recombination ($z \approx 1090$) is a  crude approximation. The angular scale of the sound horizon is a ratio of the size of the sound horizon at decoupling to the angular diameter distance to the CMB. In other words it is a ratio of some function of the evolution of the expanding universe before the CMB  and the evolution of the universe afterwards. Therefore taking it as a fixed moment in time is imprecise. 

Here, we therefore extend this approach by performing a Markov Chain Monte Carlo analysis of cosmological parameters. We allow four important parameters to vary ($\Omegam , \Omega_{b}h^{2}, h$ and $\Gamma$) for a range of set values of $\epsilon$. This allows us not to have to assume some initial fixed amount of dark matter, instead, we only demand that the amount of dark matter that has decayed in the very early universe ($z<1210$) is negligible. Furthermore, by not basing assumptions on the results of CMB experiments, the CMB measurements can be used to constrain the models.   

We conduct the analysis by using the publicly available code provided by \citet{Aubourg:2014aa}\footnote{https://github.com/slosar/april}. The data we use here is the same as in \cite{Aubourg:2014aa} and are the Galaxy BAO measurements \cite{Anderson:2014aa,Beutler:2011aa,Ross:2015aa}, BOSS Lyman-$\alpha$ forest BAO \cite{Delubac:2015aa,Font-Ribera:2014aa},  CMB data from the Planck14 results \cite{Planck-Collaboration:2014aa} combined with low-l WMAP polarization \cite{Hinshaw:2013aa} and the 740 type Ia supernovae of the Joint Light-curve Analysis \cite{Betoule:2014aa}. In the latter there is not only an increased number of supernovae compared to the Union2.1 catalog \cite{Suzuki:2012aa} used in \cite{Blackadder:2014aa} but there is also a better understanding of systematic uncertainty and calibration \cite{Aubourg:2014aa}.

We will now discuss how each one of these cosmological probes constrains the physics of decaying dark matter. 

\subsection{Baryon Acoustic Oscillations} 

Baryon Acoustic Oscillations define a cosmological standard ruler \cite{Blake:2003aa, Hu:2003aa, Eisenstein:2005aa,Eisenstein:2007aa,Bassett:2010aa}. In the early universe, baryonic matter and photons are very tightly coupled. If there is a density perturbation then, as a result of the high temperature of the photons, there will also be a pressure perturbation. The resulting sound wave expands until the end of drag epoch, $z_{d}$, after which baryons are no longer affected by compton drag and gravitational instability dominates \cite{Hu:1996aa}. This sets a fixed physical scale, a standard ruler \cite{Weinberg:2013aa}

\begin{equation}
r_{d} = \int_{z_{d}}^{\infty} \frac{c_{s}(z)}{H(z)}dz
\end{equation}
where the speed of sound is $c_{s}(z) = 3^{-1/2}c[1 + \frac{3}{4}\rho_{b}(z)/\rho_{\gamma}(z)]^{-1/2}$. After $z_{d}$, the over density will only continue to expand with the expanding universe.

The over density causes galaxies to be formed preferentially along the edges of spheres with radius $\sim r_{d}$. This can be seen in a small but characteristic peak in the correlation function of the comoving separation between galaxies \cite{Eisenstein:2005aa}.

The observed position of this peak is usually quoted as a ratio of $D_{V}/r_{d}$ where 
\begin{equation}
D_{V}=\left[(1+z)^{2}D_{A}^{2}(z)\frac{cz}{H(z)}\right]^{1/3}
\end{equation}
which is the cube root of a volume measurement that comprises the square of transverse distance (where $D_{A}$ is the angular diameter distance) and radial distance \cite{Addison:2013aa}.

Interestingly though, results given in this form are model dependent. This is because the comoving separation between two points cannot be directly measured and instead must be inferred from the redshift which requires assuming a particular cosmological model. 

The difference between two models is characterized by the factors $\alpha_{t}$ and $\alpha_{r}$ which are the fractional differences in the transverse and radial directions respectively. These factors are defined as \cite{Busca:2013aa}

\begin{eqnarray}
\alpha_{t} &\equiv& \frac{D_{A}(z)/r_{d}}{D_{A,f}(z)/r_{d,f}} \\
\alpha_{r} &\equiv& \frac{H_{f}(z)r_{d,f}}{H(z)r_{d}}
\end{eqnarray}
where $f$ denotes the fiducial model originally used to calculate the comoving separation. 

The correlation function can be expressed as a function of these two factors \cite{Xu:2013aa}. For different combinations of these factors, a $\chi^{2}$ value can be calculated based on how well the calculated correlation function fits the observations. Thus the new model can be constrained.

The BOSS results in \cite{Anderson:2014aa} comprise an analysis of observations from the Sloan Digital Sky Survey (SDSS) over an area of 8377 deg$^{2}$. \citet{Anderson:2014aa} split the observations into two independent samples, BOSS CMASS and BOSS LOWZ. The CMASS sample consists of galaxies in the redshift range $0.43 < z < 0.7$ and attempts to select galaxies of approximately constant mass based on models of galaxy evolution with redshift \cite{Aubourg:2014aa}. The LOWZ sample looks for red galaxies in the range $0.15 < z < 0.43$

The results in \cite{Beutler:2011aa} come from the 6 Degree Field Galaxy Survey (6dFGS) which looks at very low redshift galaxies to give constraints at $z = 0.106$. \citet{Ross:2015aa} contains an analysis of SDSS galaxies in the redshift range $0.07 < z < 0.2$ that reconstructs the linear fluctuation based on phase information.  These results are combined with those in \cite{Anderson:2014aa} to give a large data set of galaxy BAO observations over a wide range of redshifts (again following \cite{Aubourg:2014aa}). 

An alternative method for observing BAO is found in \citet{Delubac:2015aa} and \citet{Font-Ribera:2014aa}. They look for absorption lines in the spectra of distance quasars. Specifically they look for lines corresponding to Lyman-$\alpha$ absorption by neutral hydrogen. The existence and redshift of large amounts of neutral hydrogen along the line of site to the quasar can therefore be deduced allowing the positions of large scale structure at high redshifts to be inferred. These Lyman-$\alpha$ observations give measurements of BAO in the redshift range $2.1 < z < 3.5$, well above the range possible from direct galaxy observations.

It should be noted that the results in \citet{Beutler:2011aa} and \citet{Ross:2015aa} and from the LOWZ sample of \cite{Anderson:2014aa} are small and cannot produce a robust anisotropic analysis (that is, with separate transverse and radial components $\alpha_{t}$ and $\alpha_{r}$), so instead a single component is defined, 

\begin{equation}
\alpha = \frac{D_{V}(z)/r_{d}}{D_{V,f}(z)/r_{d,f}}
\end{equation}
For this single component the assumption is that the peak in the correlation function is modeled as a gaussian \cite{Aubourg:2014aa}.

\subsection{Cosmic Microwave Background}

Constraints from the Cosmic Microwave Background (CMB) come from the published Planck 2014 release \cite{Planck-Collaboration:2014aa}. Just as the baryon acoustic oscillations result in an imprint of a preferred scale on the matter component, the CMB contains a wealth of information of the mechanics of the process of recombination in the early universe. While in the early universe baryonic matter and photons are tightly coupled to one another, as the universe expands the temperature of the universe falls to the point that allows electron and proton recombination causing photons to decouple from baryons. This occurs at redshift $z \approx 1090$\cite{Planck-Collaboration:2014aa} although, because not all electrons and protons combine instantaneously, there is naturally some uncertainty about this value \cite{Kolb:1990aa} that affects the strength of constraints found in this paper. (Note that baryons decouple from photons at the end of the drag epoch \cite{Hu:1996aa} which occurs at approximately $z_{d}\approx 1060$\cite{Planck-Collaboration:2014aa}.) These decoupled photons then propagate and redshift until today where they are microwaves of temperature $T_{\gamma}\approx 2.7$K \cite{Fixsen:2009a}. 

The latest CMB measurements come from the PLANCK experiment \cite{Planck-Collaboration:2014ac}. Between 2009 and 2013 the satellite imaged the entire sky in multiple wavelengths and at an angular resolution that encompasses most of the major features of the expected angular anisotropy power spectrum of the CMB. The first major data release was published in 2014 \cite{Planck-Collaboration:2014aa}, and it is these results  that are used here.

In keeping with the code and method employed by \cite{Aubourg:2014aa}, our models are constrained by the Planck observations of the CMB in the following way. The Planck 2014 results for the Planck low-l lowLike model (which is a combination of Planck temperature data \cite{Planck-Collaboration:2014aa} and WMAP9 polarization data \cite{Hinshaw:2013aa}) give measurements, variances, and co-variances of $\omega_{b}$, $\omega_{\mathrm M}$ and $D_{\mathrm A}(z=1090)/r_{d}$. Because we are interested in decaying dark matter, the quantities $\omega_{b}$ and $\omega_{\mathrm M}$ are constrained based on what values they would have had if there had been no decay. 

\subsection{Supernovae Type Ia}

Finally observations of type Ia supernovae are used to constrain the decaying dark matter models, in the same spirit as in \cite{Blackadder:2014aa}.  Supernovae type Ia are believed to have a luminosity determined by their light-curve and the characteristics of their host galaxy\cite{Garavini:2007aa, Sullivan:2010aa} and thus are used as cosmological standard candles. This property allows the relative radial distances between them to be deduced \cite{Betoule:2014aa}. This is expressed as the distance modulus, $\mu$, which is simply the difference between the apparent and absolute magnitude of the supernova. The constraining power comes from calculating the theoretical distance modulus for a given cosmology using the relationship between distance modulus and luminosity distance

\begin{equation}
\mu(z) = 5 \log_{10}d_{L}(z) - 5
\end{equation}
where
\begin{eqnarray}
d_{L}(z) &=& \frac{c(1+z)}{H_{0}}\int_{0}^{z}{\cal F}^{-1/2}(z^{\prime})dz^{\prime} \\
{\cal F}(z^{\prime}) &\equiv& \Omega_{0}(z^{\prime}) + \Omega_{1}(z^{\prime}) + \Omega_{2}(z^{\prime}) \nonumber\\&&+\, \Omega_{\nu}(z^{\prime}) + \Omega_{\gamma}(z^{\prime}) + \Omega_{b}(z^{\prime}) + \Omega_{\Lambda}
\end{eqnarray}¥

The analysis in this paper (as in \cite{Aubourg:2014aa}) uses the compressed representation of the Joint Light-curve Analysis results \cite{Betoule:2014aa} of 740 supernovae. The expected distance modulus for thirty logarithmically spaced redshift bins between $z=0.01$ up to $1.3$ are calculated and then compared to the observed distance moduli. A $\chi^{2}$ value is found for each bin and then these values are summed together.

\begin{figure*}[htbp]
\begin{center}
\includegraphics[width=0.495\textwidth]{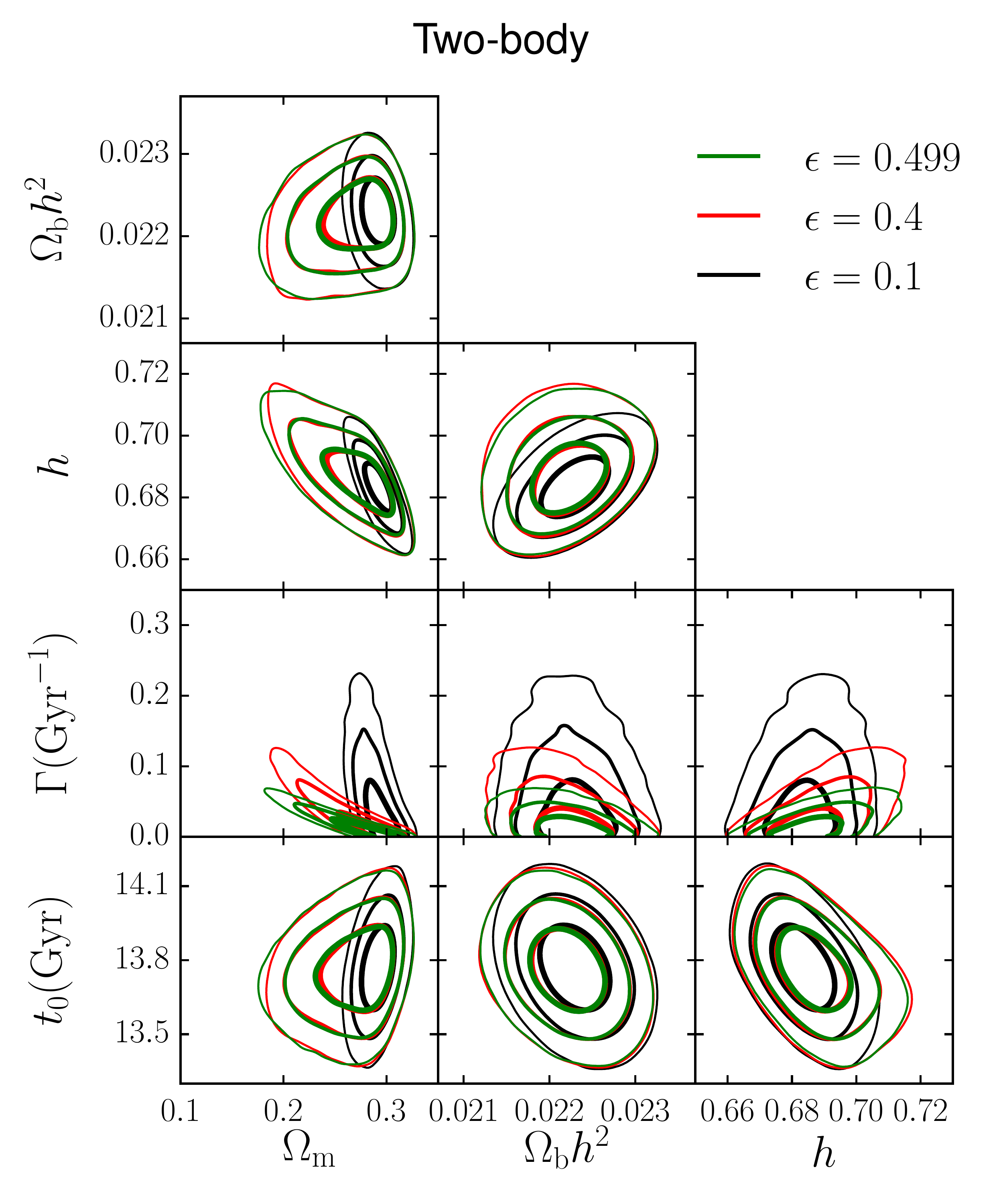} 
\includegraphics[width=0.495\textwidth]{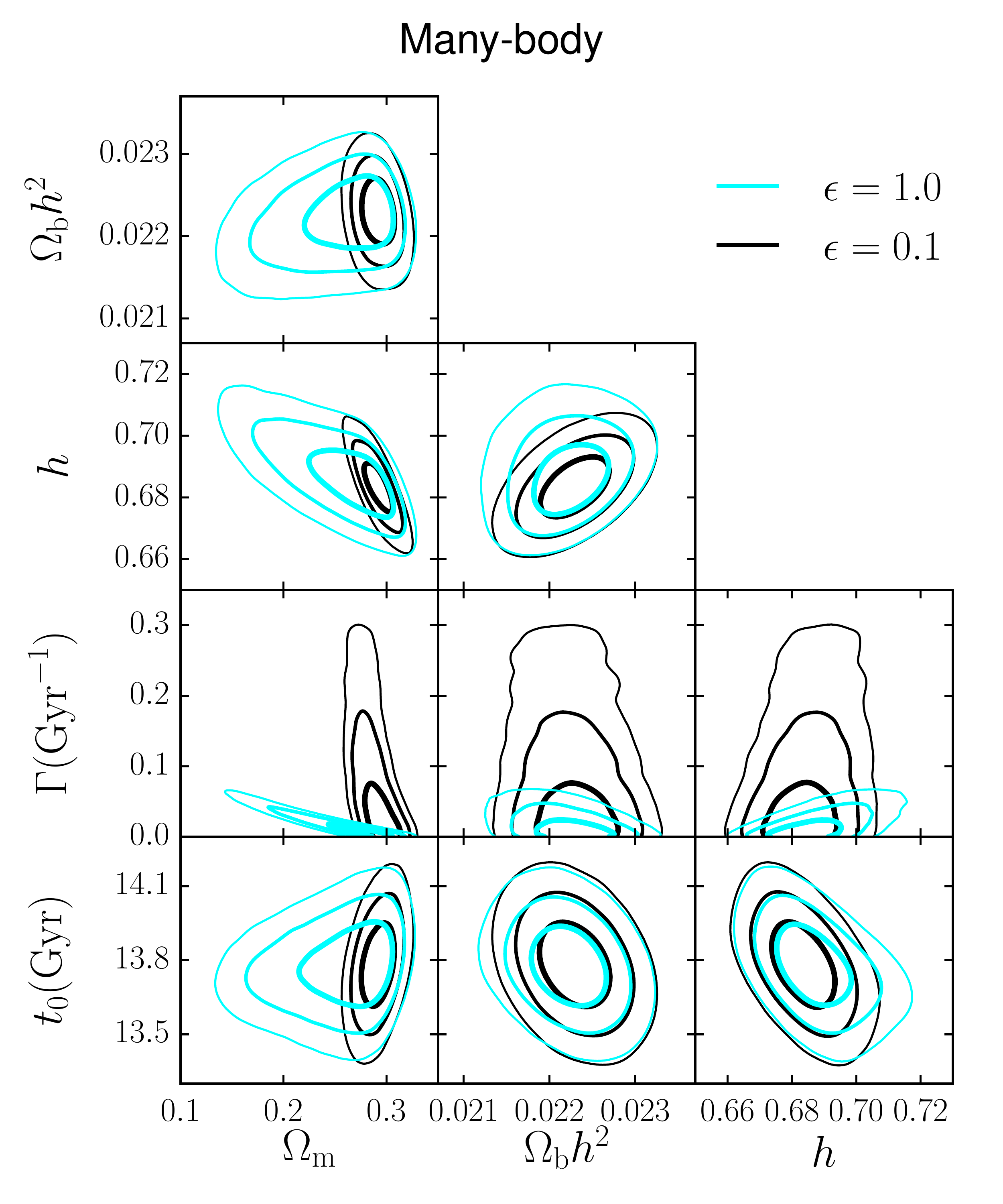} 
\caption{Results of the joint analysis of cosmological probes in the context of two-body decay (left) and many-body decay (right). The contours depict the $68\%$ (thickest line), $95\%$ and  $99.7\%$ (thinest line) confidence, while colors correspond to different values of the fraction $\epsilon$ of the parent particles' energy that is transferred to the daughter particle(s) as shown in the legend.}
\label{fig:matrix}
\end{center}
\end{figure*}
\section{\label{sec:level4} Results} 

\begin{figure}[tbp]
\begin{center}
\includegraphics[width=0.4\textwidth]{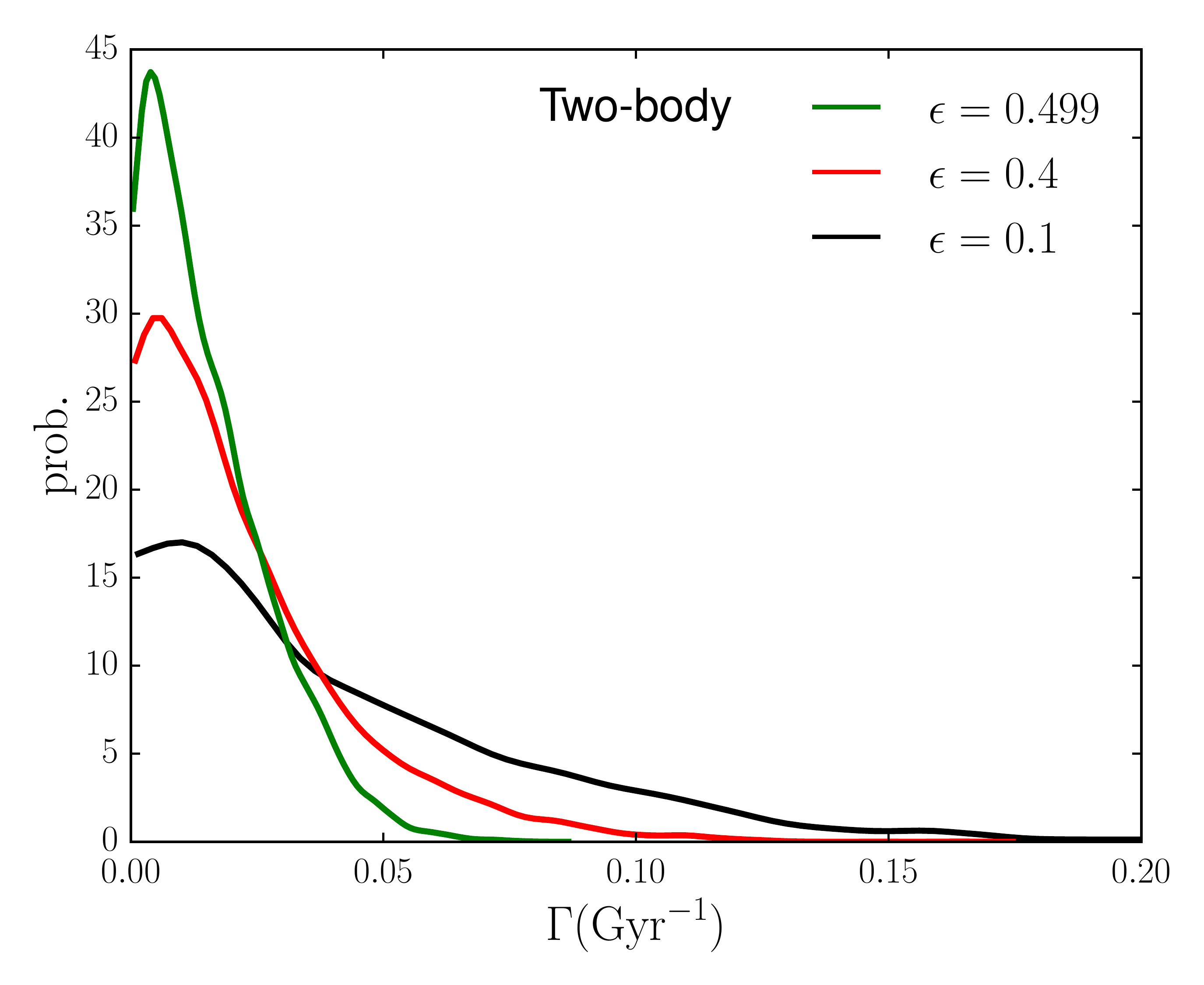} \\
\includegraphics[width=0.4\textwidth]{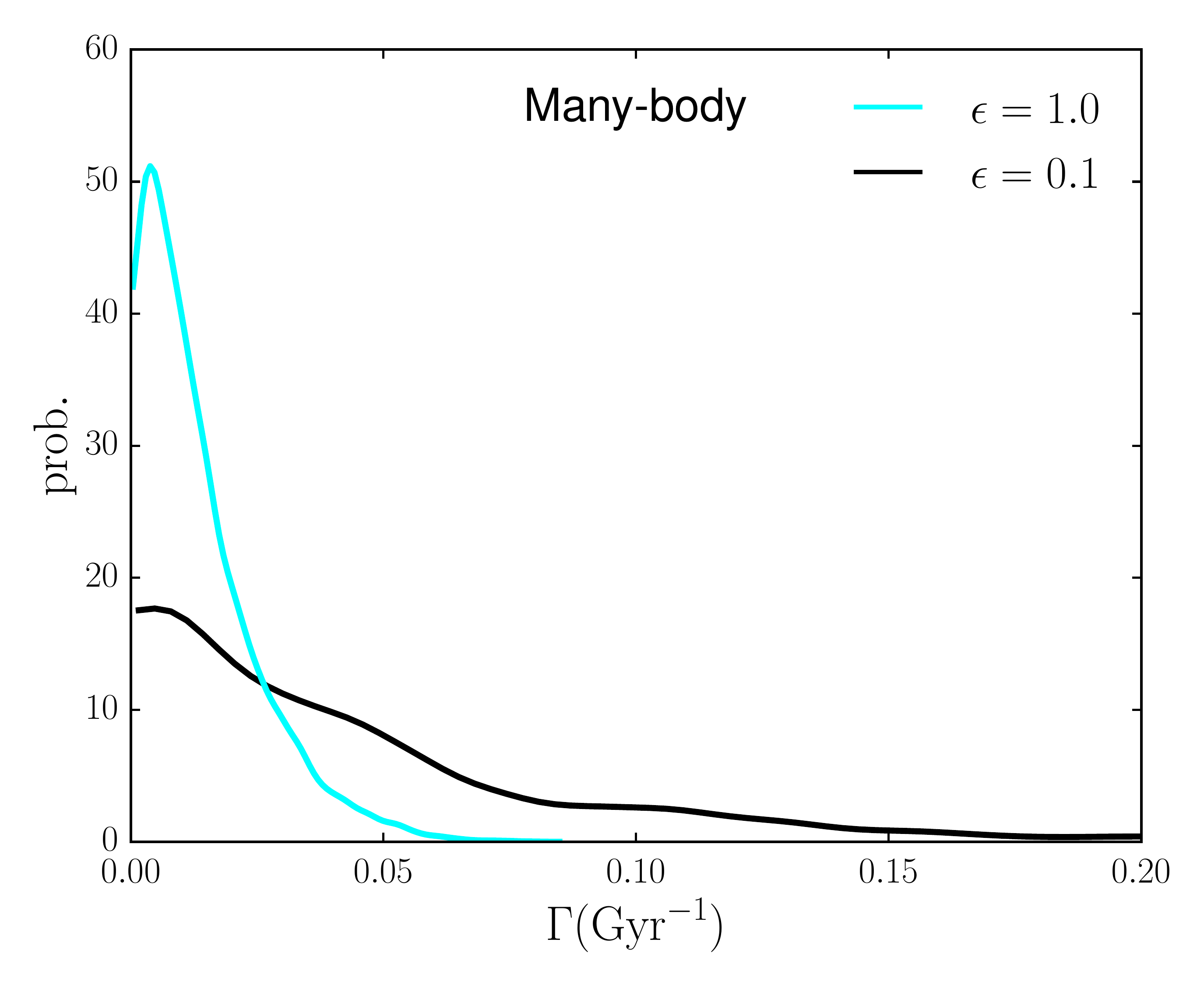} 
\caption{Probability distribution functions of the decay rate $\Gamma$ for two-body decay (top) and many-body decay (bottom). Table~\ref{tab:gamma} shows the derived $95\%$ confidence limits on the decay rate for various values of the parent particles' energy transfer, $\epsilon$, to the massless daughter particle(s).}
\label{fig:Gamma-1}
\end{center}
\end{figure}

We performed a Markov Chain Monte Carlo analysis that allowed $\Omegam$, $\Omega_{\mathrm{b}} h^2$, $h$ and $\Gamma$ to vary. This was done for a number of different values of $\epsilon$. The combined constraints in this 4 parameter space are given in the matrix plots of figure \ref{fig:matrix} and the marginal constraints on the decay rate are given in figure \ref{fig:Gamma-1}.

Fig.~\ref{fig:matrix} summarizes the results of cosmological constraints to the different decaying dark matter scenarios explored in this paper. The thick to thin contours correspond to 68\%, 95\% and 99.7\% confidence intervals respectively, while the color of contours corresponds to different values of $\epsilon$ as shown in each figure. 

Every plot in the contour matrixes exhibits overlap between the different $\epsilon$ values.  Clearly when the lifetime is long, tending towards infinity, all the contours should overlap as each model approximates the  $\Lambda$CDM model. The fact that the overlap occurs over the region of $\epsilon=0.1$ is to be expected as small epsilon values more closely resemble $\Lambda$CDM for all decay rates. 

This leads to a second observation that for larger $\epsilon$ values there is more spread in the allowed values of $\Omegam$. This is most clearly illustrated in the plot of $\Omegam$ vs $\Gamma$. It shows a dramatic effect on the allowed dark matter density for increasing decay rates. Indeed for $\epsilon=1$ in the many-body, and $\epsilon=0.499$ in the two-body, there is an almost linear relationship between an increasing decay rate and a decreasing matter density. In a very straightforward way, this plot shows that rapid dark matter decay decreases the present dark matter density (as one would expect). 

However the reduction in $\Omegam$ cannot be put down solely to the direct effect of decay. There is a slight downward trend in the allowed values of the physical baryon density, $\Omega_{b}h^{2}$, for smaller allowed values of $\Omegam$. The column of $\Omega_{b}h^{2}$ plots indicates that smaller values of the physical baryon density correspond to larger values of the decay rate and also to larger values of the present rate of the universes expansion, $h$. Cosmologies with larger decay rates still fit the data if $h$ is large. These cosmologies have larger dark energy densities arising in part due to a smaller amount of dark matter (consequently reducing the number of decays)and partly due to a reduced baryon density. Furthermore these cosmologies have had, in their recent history, more radiation because of the decay, although redshifting has a damping affect on radiation density. The faster expansion rate is a result of this greater dark energy density. 

Interestingly, we find that the age of the universe is relatively unaffected by the presence of decaying dark matter. This is naively in contrast to the apparent increase in the present value of the hubble parameter $h$, where for large values of $\epsilon$ we see an anticorrelation with the matter density. However, it should be noted that the constancy of the age of the universe is due to the fact that it is obtained through an integral of the expansion history of the universe, and thus even though decaying dark matter allows values of the present value of $h$ to vary, the age of the universe remains relatively unaffected. Cosmologies with large amounts of radiation produced by decays also have larger dark energy densities. These two effectively counterbalance leaving the present age of the universe unaffected. 




Now lets look at how the many-body and two-body matrix of plots differ. While the two models share many of the same features, the differences in the two decaying scenarios are the values of $\epsilon$ at which they occur. The two-body $\epsilon=0.4$ contours extend slightly further into small values of $\Omegam $ than the many-body $\epsilon=0.5$ This could indicate that the very elevated equation of state that occurs for the massive daughter particles in the two-body model at this $\epsilon$  is large enough to have an effect on the evolution of the universe. However for the largest plotted values of $\epsilon$ in the two-body decay model the interesting feature is that, even for $\epsilon=0.499$, the allowed region does not extend as low in $\Omegam $ compared to many-body decay. This may initially seem curious as the two-body body decay where $\epsilon$ tends towards $0.5$ creates two relativistic particles, one with zero mass, the other having mass tending towards zero which seems fairly close to the many-body decay model with $\epsilon=1$. However because one of the particles does have mass it experiences a different dynamical evolution (different equation of state) from a massless particle. Specifically $\omega_{2}$ falls towards zero and at some point the rest mass dominates the energy of the particle. It is this property that prevents lower values of $\Omegam $. 

Figure~\ref{fig:Gamma-1} shows the probability distribution just of $\Gamma$. The limit at $95\%$ confidence for the values of $\Gamma$ are given in Table \ref{tab:gamma}. Note that it is standard in the literature to give the limits in terms of $\Gamma^{-1}=\tau$. To calculate the final column, $\tau$ and $t_{0}$ were calculated at each step in the MCMC. $\tau$ was simply equal to $\Gamma^{-1}$ while $t_{0}$ was found by taking the integral of $[aH(a)]^{-1}$ with respect to $a$. 

\begin{table}[]
\centering
\caption{$95\%$ Confidence Limits}
\label{tab:gamma}
\begin{tabular}{|c|c|c|c|c|}
\hline
model & $\epsilon$ & \begin{tabular}[c]{@{}c@{}}$\Gamma$ (Gyr$^{-1}$)\\ Upper Limit\end{tabular} & \begin{tabular}[c]{@{}c@{}}$\Gamma^{-1}$ (Gyr)\\ Lower Limit\end{tabular} & \begin{tabular}[c]{@{}c@{}}$\tau/t_{0}$\\ Lower Limit\end{tabular} \\ \hline
Two   & 0.499      & $0.040$                                                                                              & $25$                                                                                               & $1.8$                                                                                       \\
Two   & 0.49       & $0.045$                                                                                              & $22$                                                                                               & $1.6$                                                                                       \\
Two   & 0.45       & $0.054$                                                                                              & $19$                                                                                               & $1.4$                                                                                       \\
Two   & 0.4        & $0.067$                                                                                              & $15$                                                                                               & $1.1$                                                                                       \\
Two   & 0.3        & $0.074$                                                                                              & $13$                                                                                               & $0.98$                                                                                      \\
Two   & 0.2        & $0.12$                                                                                               & $8.4$                                                                                              & $0.61$                                                                                      \\
Two   & 0.1        & $0.12$                                                                                               & $8.4$                                                                                              & $0.61$                                                                                      \\
Many  & 1          & $0.037$                                                                                              & $27$                                                                                               & $2.0$                                                                                       \\
Many  & 0.5        & $0.069$                                                                                              & $14$                                                                                               & $1.1$                                                                                       \\
Many  & 0.1        & $0.15$                                                                                               & $6.7$                                                                                              & $0.48$                                                                                      \\ \hline
\end{tabular}
\end{table}

\section{\label{sec:level5}Discussion}

There has been rather interesting work carried out in decaying dark matter since the publication of \cite{Blackadder:2014aa}. Here we discuss the relevance of this work to the rest of the literature. 

After the apparent discovery of a $3.5$keV line emission from Andromeda and Perseus \cite{Boyarsky:2014aa}, there have been hypotheses that claim that its presence has a decaying dark matter origin.  \citet{Lovell:2014aa} find consistency between observations of what seems to be an X-ray excess from the galactic center \cite{Daylan:2014aa,Hooper:2011aa} and from Andromeda by assuming dark matter of mass $\approx 7.1$keV decaying in a two-body decay to photons with a lifetime of $\sim 10^{28}$s or $\sim 3 \times 10^{11}$Gyr. (For a review of the $3.5$keV emission line and other possible explanations see \citet{Iakubovskyi:2014aa}.)

Several papers considered models equivalent to many-body decaying dark matter with $\epsilon=1.$ \citet{Audren:2014aa} performed an MCMC using Planck14 low-l, high-l and lensing reconstruction \cite{Planck-Collaboration:2014aa}, WMAP polarization \cite{Hinshaw:2013aa}, Wiggle Z \cite{Parkinson:2012aa} and the BOSS BAO measurement at redshift $z=0.57$ {\cite{Anderson:2014aa}. They found a lower limit on the dark matter lifetime of $160$Gyr at $95\%$ confidence.

\citet{Enqvist:2015aa} also looked at $\epsilon=1$ many-body decays. They employed N-body simulations to find non-linear corrections to the matter power spectrum. Interestingly when they constrained their model against CMB data alone (Planck14 temperature maps \cite{Planck-Collaboration:2014ab} and WMAP9 polarization \cite{Bennett:2013aa}) they found a lower limit at $95\%$ confidence of $140$Gyr, but when comparing to a combination of CMB and weak lensing data \cite{Heymans:2013aa} the limit was reduced to $97$Gyr. Based on this \citet{Enqvist:2015aa} posited that decaying dark matter was relieving the tension in the measurement of the amplitude of matter fluctuations $\sigma_{8}$ that exists between the different observations. 

By considering that the daughter products of dark matter decay might ionize and heat the interstellar medium, \citet{Yang:2015aa} found a $95\%$ lower limit on the lifetime as $\approx 1.3 \times 10^{9}$Gyr. However to achieve this limit the author assumed that $f(z)$, the fraction of decay energy deposited in the medium, was constant with redshift and that the constant equaled 1 (the authors did not provide a justification for these two approximations).

Models of decaying dark matter with specific decay channels have been used in attempts to solve outstanding problems. For example \citet{Geng:2015aa} found that a two-component decaying dark matter, with a heavy particle decaying to $\mu^{+}\mu^{-}$ and a lighter particle decaying to $\tau^{+}\tau^{-}$, could fit both AMS-02 \cite{Aguilar:2013aa}  and Fermi-LAT \cite{Abdo:2009aa}  data. Meanwhile \citet{Hamaguchi:2015aa} found that a decay into $W^{\pm}l^{\mp}$ could explain the AMS-02 antiproton observation. In particular this model, in the case of a graviton with R-parity violation, gave rise to a particle of mass $M\sim 1$TeV with lifetime $\sim 3 \times 10^{10}$Gyr. This is around 9 orders of magnitude stronger than our strongest constraints, but what  \citet{Hamaguchi:2015aa} gain in constraining power they lose in generality by specifying a detailed model. 

\citet{Esmaili:2014aa} have found a slight preference ($\sim 2\sigma$) in the IceCube observations  \cite{Aartsen:2014aa} of high energy neutrinos which they assert are distributed more in the manner expected of decaying dark matter than of a purely isotropic astrophysical source. To provide the neutrinos with enough energy, the decays would have to come from an ultra heavy dark matter candidate. Their paper then makes the leap of assuming that the high energy neutrinos can indeed be described by astrophysical sources and on this they then set a limit on decaying dark matter in the TeV range of having a lifetime of at least $10^{26}-10^{28}$s or $3 \times 10^{9} - 3 \times 10^{11}$Gyr. \citet{Rott:2014aa} also considered a heavy dark matter particle decaying into a neutrino and some other standard model particles. In particular they assumed that the parent had mass greater than $100$TeV (breaking the unitarity bound) and therefore  IceCube results bound the lifetime to $\tau > 3 \times 10^{11}$Gyr -- approximately 10 orders of magnitude in constraining strength has been gained at the expense of a much less generally applicable model.

If dark matter decays into standard model particles this can give rise to gamma-rays. \citet{Ando:2015aa} the authors consider decay into six standard model particle channels, motivated by beyond the standard models particles. These decay products ($\nu l^{+} l^{-}, l \bar{l}, W^{\pm}, q, \bar{q}$) add to the extra-galactic gamma-ray background, EGRB, and can be constrained using Fermi-LAT data. Over the various decay channels (and assuming other EGRB contributions from blazars, star forming galaxies and misaligned active galactic nuclei) the lifetime can be constrained to be at least  $3 \times 10^{9} - 3 \times 10^{11}$Gyr. By specifying the decay products,  \citet{Ando:2015aa}  is able to achieve a minimum of 8 orders of magnitude improvement in constraints compared to our more general model.

Clearly many of the above models are more specific than the generic models considered here and so their bounds were correspondingly much stronger. By considering generic models of decaying dark matter, in which some of the decay products are massless and relativistic, broadly applicable bounds have been obtained. This is the case in three recent papers that considered generic models of many-body decay\cite{Aubourg:2014aa,Audren:2014aa,Enqvist:2015aa}. Of those, all but \citet{Aubourg:2014aa} considered only $\epsilon=1$. Both \cite{Audren:2014aa,Enqvist:2015aa} reported much stronger bounds on the lifetime compared to this work. Both focus on how decaying dark matter affects the matter power spectrum. \citet{Audren:2014aa} applies linear corrections to the matter power spectrum and finds a lifetime of at least $160$Gyr at $95\%$ confidence, while \citet{Enqvist:2015aa} uses N-body simulations to find non-linear corrections to the matter power spectrum. Note that \citet{Enqvist:2015aa} focuses on obtaining constraints at or more than lifetimes of $100$Gyr, partly motivated by the constraints of \citet{Audren:2014aa} and the findings of an early draft of \citet{Aubourg:2014aa} (which has since been updated with considerably weaker decaying dark matter constraints). They find a $95\%$ confidence lower bound of $97$Gyr, which is nearly $40\%$ weaker than  \cite{Audren:2014aa}, however it is not clear how the assumptions of large lifetimes affects the result. 

Finally, during the final preparations of this manuscript, \citet{Aubourg:2014aa} updated their results and find that for $h=0.68$, the lifetime is constrained to be $\Gamma^{-1} > 28$ Gyr at 95\% confidence level for a many-body decay with $\epsilon =1$. This is consistent with what we find here (see Table~\ref{tab:gamma}). The small difference can be attributed to the marginalization of $h$ that we performed in this work, as compared to a fixed value of $h$ quoted in \citet{Aubourg:2014aa}.

In summary, we have taken the model derived in \cite{Blackadder:2014aa} and placed constraints based on data from the cosmic microwave background, baryon acoustic oscillations and type Ia supernovae. The constraints on two-body decay are the strongest in the literature as are those for many-body decays with $\epsilon \neq 1$. The models and results are sufficiently general as to be widely applicable to many possible dark matter candidates.

\acknowledgements
We acknowledge useful conversations with Ian Dell'Antonio, Alex Geringer-Sameth, Stefan Piperov. We especially thank An$\breve{\mathrm{z}}$e Slosar and David Weinberg for clarifications and help with the publicly available code that was used in this project. This work was supported by Brown University through the use of the facilities of its Center for Computation and Visualization. SMK partially is supported by DOE DE-SC0010010, and NSF PHYS-1417505. GB is partially supported by NSF PHYS-1417505.  

\bibliography{BK_2015.bib}
\end{document}